\newif\ifarxiv
\arxivtrue

\ifarxiv
  \documentclass{article}
\else
  \documentclass[useAMS,referee]{biom}
\fi

\usepackage[utf8]{inputenc} % allow utf-8 input
\usepackage[T1]{fontenc}    % use 8-bit T1 fonts
\usepackage{booktabs}       % professional-quality tables
\usepackage{amsfonts, comment}       % blackboard math symbols
\usepackage{nicefrac}       % compact symbols for 1/2, etc.
\usepackage{microtype}      % microtypography
\usepackage{amsmath}
\usepackage{subcaption}
\usepackage{hyperref}
\usepackage{url}
\let\bibhang\relax
\usepackage{natbib}
\usepackage{graphicx}
\usepackage{xcolor}
\usepackage[group-separator={,},group-minimum-digits=4]{siunitx}
\ifarxiv
  \usepackage{arxiv}
\else 
  
\fi

\ifarxiv
  \title{Symmetric Vaccine Efficacy}
    \author{Lucy D'Agostino McGowan$^{1*}$, Sarah C.~Lotspeich$^{1}$, and Michael G.~Hudgens$^{2}$\\
	$^{1}$Department of Statistical Sciences,
	Wake Forest University,
	Winston-Salem, NC 27106 \\
	$^{2}$Department of Biostatistics
	University of North Carolina at Chapel Hill,
	Chapel Hill, NC 27599 
    }
\else
  \title[Symmetric Vaccine Efficacy]{Symmetric Vaccine Efficacy}

  \author{Lucy D'Agostino McGowan$^{1*}$\email{mcgowald@wfu.edu}, Sarah C.~Lotspeich$^{1}$, and Michael G.~Hudgens$^{2}$\\
	$^{1}$Department of Statistical Sciences,
	Wake Forest University,
	Winston-Salem, NC 27106 \\
	$^{2}$Department of Biostatistics
	University of North Carolina at Chapel Hill,
	Chapel Hill, NC 27599 
    }
\fi

\begin{document}

\ifarxiv 

\maketitle

\begin{abstract}
Traditional measures of vaccine efficacy (VE) are inherently asymmetric, constrained above by $1$ but unbounded below. As a result, VE estimates and corresponding confidence intervals can extend far below zero, making interpretation difficult and potentially obscuring whether the apparent effect reflects true harm or simply statistical uncertainty. The proposed symmetric vaccine efficacy (SVE) is a bounded and interpretable alternative to VE that maintains desirable statistical properties while resolving these asymmetries. SVE is defined as a symmetric transformation of  infection risks, with possible values within $[-1, 1]$, providing a common scale for both beneficial and harmful vaccine effects. This paper describes the relationship between SVE and traditional VE, considers inference about SVE, and illustrates the utility of the proposed measure by reanalyzing data from a randomized trial of a candidate HIV vaccine. 
Open-source tools for computing estimates of SVE and corresponding confidence intervals are available in R through the \texttt{sve} package.\\ 
\end{abstract}


\else

\pagerange{\pageref{firstpage}--\pageref{lastpage}} 
\date{{\it Received December} 2025.}

\pagerange{\pageref{firstpage}--\pageref{lastpage}} 

\volume{XXX}
\pubyear{XXXX}
\artmonth{month}
\doi{000-000}
\label{firstpage}
\begin{abstract}
Traditional measures of vaccine efficacy (VE) are inherently asymmetric, constrained above by $1$ but unbounded below. As a result, VE estimates and corresponding confidence intervals can extend far below zero, making interpretation difficult and potentially obscuring whether the apparent effect reflects true harm or simply statistical uncertainty. The proposed symmetric vaccine efficacy (SVE) is a bounded and interpretable alternative to VE that maintains desirable statistical properties while resolving these asymmetries. SVE is defined as a symmetric transformation of  infection risks, with possible values within $[-1, 1]$, providing a common scale for both beneficial and harmful vaccine effects. This paper describes the relationship between SVE and traditional VE, considers inference about SVE, and illustrates the utility of the proposed measure by reanalyzing data from a randomized trial of a candidate HIV vaccine. 
%In the example, SVE yields more interpretable results
%confidence intervals and clearer graphical summaries
%compared to standard VE. 
Open-source tools for computing estimates of SVE and corresponding confidence intervals are available in R through the \texttt{sve} package.\\ 
\end{abstract}


\maketitle

\fi

\section{Introduction}

Vaccines are one of the most effective public health interventions for preventing infectious diseases. Quantifying vaccine effects accurately and providing interpretable estimates is critical for regulatory decisions, public health policy, and individual risk communication. The most commonly used measure of vaccine performance is \textit{vaccine efficacy} (VE), typically defined as one minus the relative risk of infection if  vaccinated versus not vaccinated \citep{halloran1991direct, who}. Throughout this paper, the term \textit{infection} is used to describe the  outcome of interest for consistency and clarity; however, the framework applies equally to any binary outcome commonly measured in vaccine trials (e.g., symptomatic disease, hospitalization, or death).

The parameter space for traditional VE is inherently asymmetric, with VE bounded above by $1$ (complete protection afforded by vaccination) but unbounded below. Consequently, VE estimates and corresponding confidence intervals can take on extreme negative values when the vaccine group experiences higher infection rates or the numbers of individuals infected in both groups are small. Such extremes have been observed in certain subgroups of HIV vaccine trials \citep{rgp1202005placebo} and   analyses of COVID-19 vaccine trials \citep{buchan2022estimated, bodner2023observed, hansen2021vaccine}. Asymmetry in  VE complicates  interpretation, particularly when communicating results in settings with small numbers of infections where confidence intervals can be wide.

Symmetrization of effect measures has been proposed in other contexts to improve interpretability by placing beneficial and harmful treatment effects on a common scale. For example, \citet{berry2006symmetrized} introduced the \textit{symmetrized percent change} for treatment comparisons from baseline, demonstrating improved interpretability without loss of statistical efficiency. More broadly,  effect measures that switch form depending on the direction of the treatment effect have appeared in the statistical literature. For example, \citet{van2007estimation} proposed the switch causal relative risk for binary outcomes in the context of randomized trials with non-compliance, and \citet{baker2018new} proposed the  generalized relative risk reduction in the context of meta-analyses. \citet{huitfeldt2021shall} showed that such effect measures have desirable stability properties. This paper considers a similar approach to vaccine evaluation, defining a \textit{symmetric vaccine efficacy} (SVE) metric that provides a bounded, interpretable measure of vaccine effect ranging from $-1$ (maximal harm) to $1$ (maximal protection), and developing inferential methods including profile likelihood and delta method confidence intervals that have not previously been studied for this class of parameters.

The outline of the remainder of this paper is as follows. Section~\ref{sec:sve} compares the proposed SVE estimand to the traditional VE parameter. Section~\ref{sec:inference} considers inference about the SVE estimand. Section~\ref{sec:sim} examines operating characteristics of the inferential methods through simulation. Section~\ref{sec:data} illustrates the approach using data from a large HIV vaccine trial \citep{rgp1202005placebo}, demonstrating the interpretational advantages of SVE over traditional VE.  Section~\ref{sec:r} describes R code for estimating SVE and corresponding confidence intervals.

\section{Symmetric Vaccine Efficacy} \label{sec:sve}
VE is the standard measure used to quantify the effect of a vaccine \citep{who}. In practice, VE is often defined as one minus a relative measure of risk, such as the relative risk or hazard ratio. For example, in a trial setting with infection probabilities $p_1$ if vaccinated and $p_0$ if not vaccinated, the VE parameter is often defined as
\begin{equation}
\text{VE}=1-\frac{p_1}{p_0},
\label{eq:ve}
\end{equation} 
with $\text{VE}=1$ corresponding to complete protection (i.e., no risk of infection if vaccinated) 
and
$\text{VE}=0$ corresponding to the null of no effect (i.e., equal risk regardless of whether vaccinated)
\citep{halloran1991direct}.
The VE parameter as defined in \eqref{eq:ve} is bounded above by $1$ but unbounded below, making it inherently asymmetric. 
%In practice, VE is estimated by substituting sample proportions $\hat{p}_1$ and $\hat{p}_0$ for the true probabilities in Equation~\ref{eq:ve}, yielding $\widehat{\text{VE}}=1-\hat{p}_1/\hat{p}_0$. The estimated VE inherits the asymmetry of the parameter but can exhibit more extreme behavior due to sampling variability. When the baseline risk of infection is low (i.e., $p_0$ is small) or sample sizes are small, confidence intervals for $\widehat{\text{VE}}$ can extend to large negative values, even when the true VE is positive. 

Alternatively, a symmetric measure of vaccine efficacy (SVE) can be defined as
\begin{equation}
\text{SVE}=\frac{p_0-p_1}{\max(p_0, p_1)}.\label{eq:sve}
\end{equation}
By construction, SVE is bounded between $-1$ and $1$, and has a straightforward interpretation as follows.
\begin{itemize}
\setlength{\itemindent}{1em} % Adjust this value as needed
    \item[(i)] When $p_0>p_1$, $\text{SVE}=1-p_1/p_0$ will be greater than $0$ and equal to VE. Thus, in this case, SVE has the same interpretation as the traditional VE metric, i.e.,  the proportional reduction in risk due to vaccination. SVE values closer to $1$ indicate a stronger protective effect. 
    \item[(ii)] When $p_1>p_0$, $\text{SVE}=p_0/p_1-1$ will be less than zero. Negative SVE captures the proportional decrease in risk from not being vaccinated, with values closer to $-1$ corresponding to more possible  harm due to vaccination on the same scale as the protective effect of  vaccination.
\end{itemize}

\begin{figure}
    \centering
    \begin{subfigure}[t]{0.48\textwidth}
        \centering
        \includegraphics[width=\linewidth]{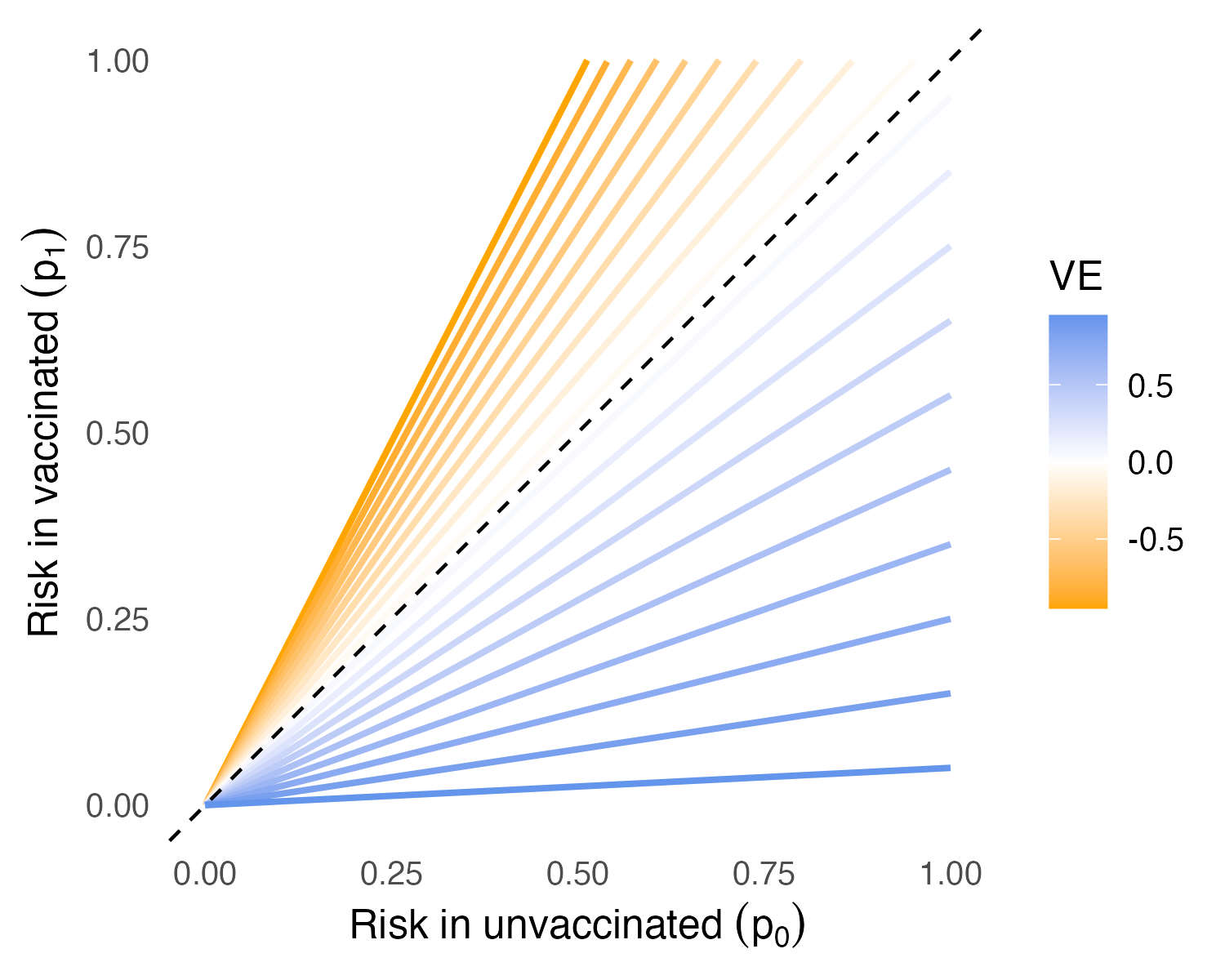}
        \caption{Traditional Vaccine Efficacy (VE)}
        \label{fig:labbe_ve}
    \end{subfigure}
    \hfill
    \begin{subfigure}[t]{0.48\textwidth}
        \centering
        \includegraphics[width=\linewidth]{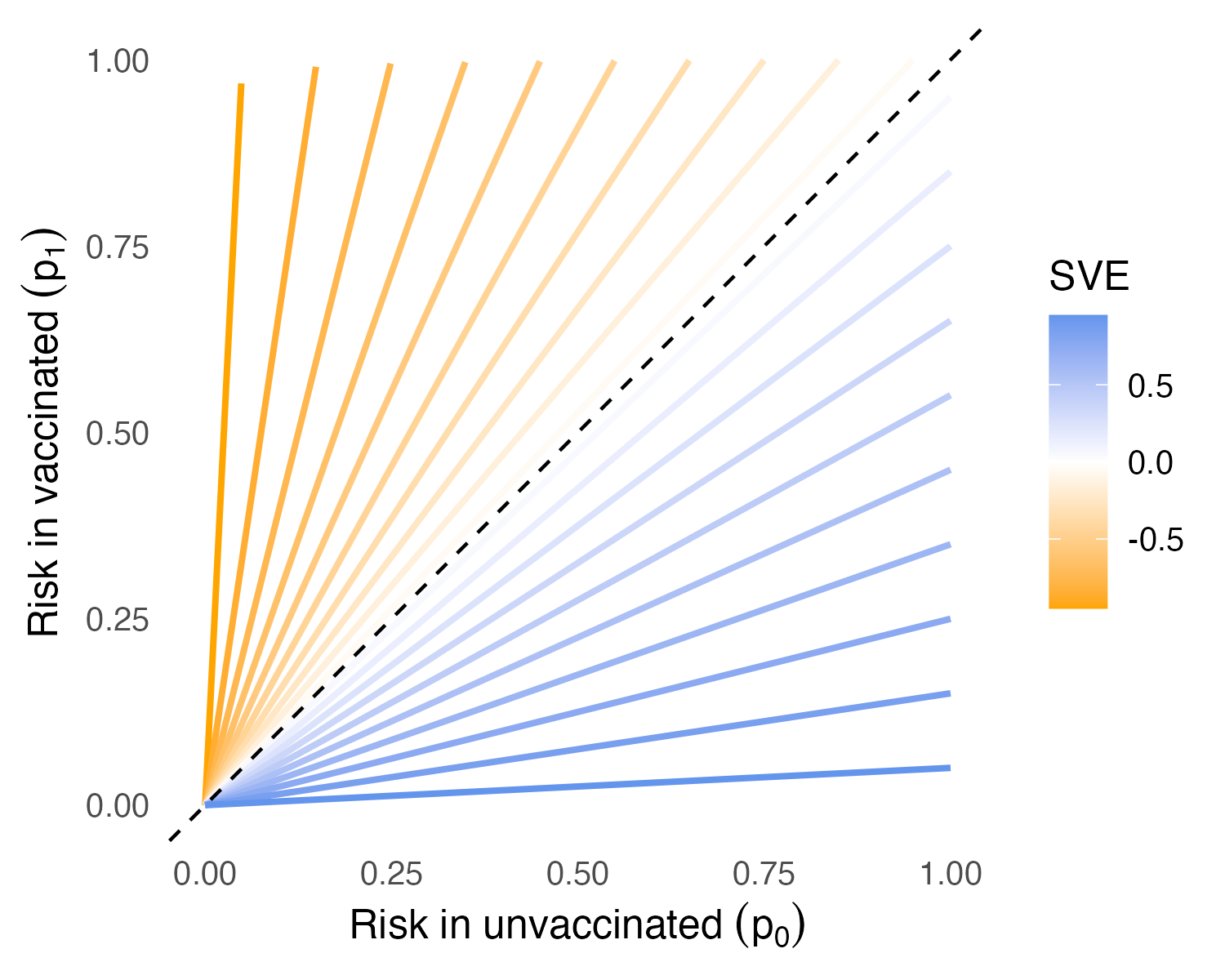}
        \caption{Symmetric Vaccine Efficacy (SVE)}
        \label{fig:labbe_sve}
    \end{subfigure}
    \caption{L'Abbe plots illustrating iso-effect lines for \textbf{(a)} traditional VE and \textbf{(b)} SVE. Each line represents a fixed effect size, with $20$ evenly spaced values from $-0.95$ to $0.95$ ranging from strong harm to strong protection represented by the line color. The dashed diagonal line at $y = x$ indicates equal risk in vaccinated and unvaccinated groups. For traditional VE, the iso-effect lines above this diagonal (negative effect sizes) appear compressed compared to those below it (positive effect sizes), illustrating the metric's inherent asymmetry. By contrast, the symmetric SVE iso-effect lines remain bounded and mirror protective and harmful effects around the diagonal, providing a more interpretable measure of vaccine efficacy.}
    \label{fig:labbe}
\end{figure}

To illustrate the contrast between traditional VE and the proposed SVE, Figure~\ref{fig:labbe} shows L'Abbe plots \citep{l1987meta} of infection probabilities if vaccinated versus if not vaccinated. In these plots, iso-effect lines correspond to pairs of infection  probabilities $(p_0,p_1)$ with the same effect size. For example, 
the $45$-degree line indicates pairs of equal infection probabilities (i.e., the vaccine has no effect such that infection risk is the same in both groups).
Iso-effect lines for traditional VE shown in Figure~\ref{fig:labbe_ve} correspond to $p_1 = (1-\text{VE})p_0$,
whereas the iso-effect lines for SVE shown in 
Figure~\ref{fig:labbe_sve} are given by $p_1 =(1-\text{SVE})p_0$ if $\text{SVE}\geq0$ and $p_1=p_0/(1+\text{SVE})$ if $\text{SVE}<0$.
In contrast to VE, the SVE iso-effect lines are symmetric about the $45$-degree line, with SVE bounded between $-1$ and $1$ for all possible  values of $(p_0,p_1)$. That is, SVE is a balanced metric where both protection and harm are represented on the same bounded scale, highlighting its interpretability advantages.

\section{Inference for Symmetric Vaccine Efficacy} \label{sec:inference}

Consider a randomized trial with perfect compliance and no loss to follow-up. Let $X_0\sim \text{Binomial}(n_0,p_0)$ and $X_1 \sim \text{Binomial}(n_1,p_1)$ denote the numbers of infections observed in the unvaccinated and vaccinated groups, respectively, where $n_0$ and $n_1$ are the group sample sizes \citep{halloran1991direct}. Assume $p_0,p_1 \in (0,1)$.
Denote the empirical risks in the two groups by $\hat{p}_0=X_0/n_0$ and $\hat{p}_1=X_1/n_1$, which are unbiased estimators of the true infection risks $p_0$ and $p_1$. Define the plug-in estimator of  SVE to be $\widehat{\text{SVE}}=(\hat{p}_0-\hat{p}_1)/\max(\hat{p}_0,\hat{p}_1)$, which is obtained by substituting the empirical risks for the true risks in \eqref{eq:sve}. By the law of large numbers and continuous mapping theorem, it follows that $\widehat{\text{SVE}}$ is a consistent estimator of SVE as $n_0,n_1 \to \infty$. 
Like other ratio estimators \citep{jewell1986bias}, 
$\widehat{\text{SVE}}$ will in general exhibit finite-sample bias.
In Web Appendix B.1
a bias-corrected estimator is proposed which may be preferred
in small sample size settings.
In Sections~\ref{subsec:profile} and \ref{subsec:wald}, confidence intervals (CIs) for SVE using the profile likelihood and delta method are presented.

\subsection{Profile Likelihood Confidence Intervals} \label{subsec:profile}

A confidence interval for SVE can be obtained by inverting a likelihood ratio test, as follows. First note that 
$\hat{p}_0$ and $\hat{p}_1$ are the unconstrained maximum likelihood estimators (MLEs) of the log-likelihood 
\begin{equation}
    \ell(p_0,p_1)=\sum_{j=0}^1\{ X_j\log(p_j)+(n_j-X_j)\log(1-p_j)\}.
    \label{eq:unconst}
\end{equation}
Likelihood-based inference of the SVE parameter can be achieved via the profile likelihood, wherein for any value of SVE the likelihood is maximized over $p_0$ and $p_1$, subject to the functional constraints implied by the value of SVE. 
In particular,   if $\text{SVE}\ge 0$,  then the constraint is $p_1=p_0(1-\text{SVE})$. On the other hand, if $\text{SVE}<0$, then  the constraint is $p_0=p_1(1+\text{SVE})$. Therefore the  profile log-likelihood for SVE is
\begin{equation}
    \ell_p(\text{SVE})=
    \begin{cases}
        \displaystyle \sup_{p_0\in (0,1)}\ell(p_0, p_0(1-\text{SVE})) & \text{if SVE}\in[0,1),\\[5pt]
        \displaystyle \sup_{p_1\in(0,1)}\ell(p_1(1+\text{SVE}),p_1) &\text{if SVE}\in(-1,0).   
    \end{cases}
    \label{eq:profile}
\end{equation}
The two cases in \eqref{eq:profile} each entail a one-dimensional maximization over either $p_0$ or $p_1$ which can be easily solved using standard one-dimensional numerical optimization algorithms.

Consider testing the null hypothesis $H_0:\text{SVE}=s$ for some $s \in (-1,1)$. Define the likelihood ratio statistic $\Lambda(s)=2\{\ell(\hat{p}_0,\hat{p}_1)-\ell_p(s)\}$, where $\ell(\hat{p}_0,\hat{p}_1)$ is the maximum of the unconstrained log-likelihood \eqref{eq:unconst}, and $\ell_p(s)$ is the profile log-likelihood  \eqref{eq:profile} evaluated at $\text{SVE}=s$. Larger values of $\Lambda(s)$ provide stronger evidence against $H_0:\text{SVE}=s$. 

Under standard regularity conditions as $n_0,n_1\to\infty$, $\Lambda(s)$ converges in distribution to a $\chi_1^2$ under the null. Using this asymptotic approximation, the likelihood ratio test can be inverted in the usual fashion to construct a CI for SVE. Specifically, the $(1-\alpha)\%$ profile likelihood CI for SVE is $\{s \in (-1,1):\Lambda(s)\leq \chi^2_{1,1-\alpha}\}$, where $\chi^2_{1,1-\alpha}$ denotes a $(1-\alpha)$-quantile of a $\chi^2_1$ distribution.

\subsection{Wald Confidence Intervals}\label{subsec:wald}

For large samples, Wald confidence intervals provide a computationally efficient alternative to the profile likelihood approach. Let $n=n_0+n_1$ be the total trial size and assume $n_0/n\rightarrow\lambda\in(0,1)$ as $n \rightarrow \infty$. Then, under standard regularity conditions and assuming independence between groups, by the bivariate central limit theorem
\begin{equation*}
\sqrt{n}
\begin{pmatrix}
\hat{p}_0-p_0\\
\hat{p}_1-p_1\end{pmatrix}\xrightarrow{d} \boldsymbol{\mathcal{N}}\left(\mathbf{0},\boldsymbol\Sigma\right),\text{where }\boldsymbol\Sigma = \begin{pmatrix}
p_0(1-p_0)/\lambda & 0\\ 
0 & p_1(1-p_1)/(1-\lambda)
\end{pmatrix}.
\end{equation*}
The multivariate delta method in turn implies that $\widehat{\text{SVE}}$ is also asymptotically normal.
In particular, let $g(p_0,p_1)=\text{SVE}=(p_0-p_1)/\max(p_0,p_1)$ such that $g(\hat{p}_0,\hat{p}_1)=\widehat{\text{SVE}}$. Note that the gradient $\boldsymbol\nabla g$ exists and is continuous throughout the interior of the parameter space, despite the maximum operator in the denominator (see Web Appendix A).
Therefore,  $\sqrt{n}(\widehat{\text{SVE}}-\text{SVE})
\xrightarrow{d} \boldsymbol{\mathcal{N}}(0,\boldsymbol\nabla g^\top\boldsymbol\Sigma\boldsymbol\nabla g)$. 
The asymptotic variance can be consistently estimated by
replacing 
$p_0$ and $p_1$
with 
$\hat p_0$ and $\hat p_1$,
and $\lambda$ with $\hat \lambda = n_1/n$. 
Letting $\hat{\sigma}_0^2=\hat{p}_0(1-\hat{p}_0)/n_0$ and $\hat{\sigma}_1^2=\hat{p}_1(1-\hat{p}_1)/n_1$, it follows
(see Web Appendix A for details)
that in large samples, the variance of $\widehat{\text{SVE}}$
can be approximated by
\begin{equation}
    \widehat{\text{Var}}(\widehat{\text{SVE}})=\frac{\hat{p}_1^2\hat\sigma^2_0+\hat{p}^2_0\hat{\sigma}^2_1}{\{\max(\hat{p}_0, \hat{p}_1)\}^4},
    \label{eq:var}
\end{equation}
which simplifies to $(\hat\sigma^2_0+\hat{\sigma}^2_1)/\hat{p}^2$ when $\hat{p}_0=\hat{p}_1=\hat{p}$.
Thus a $100(1-\alpha)\%$ Wald-type CI for SVE is $\widehat{\text{SVE}} \pm z_{1-\alpha/2} \widehat{\text{Var}}(\widehat{\text{SVE}})^{1/2}$, where $z_{1-\alpha/2}$ is the ($1-\alpha/2$) quantile of the standard normal distribution. These CIs are asymptotically valid but can include values outside the parameter space $[-1,1]$ when the SVE estimate lies near the boundaries or in settings where there is considerable uncertainty (e.g., in small samples).

Wald CIs may also be computed on a transformed scale to ensure that the lower and upper endpoints of the CI remain within the parameter space $[-1,1]$ after back-transformation. For example, the SVE estimator can be transformed using the inverse hyperbolic tangent $\hat u=\tanh^{-1}(\widehat{\text{SVE}})=0.5\log\{(1+\widehat{\text{SVE}})/(1-\widehat{\text{SVE}})\}$, where $\hat u$ can take on any real value in $(-\infty, \infty)$  
\citep{fisher1921probable,hotelling1953new, van2000asymptotic}. 
By the delta method, $\text{Var}(\widehat u) \approx  \text{Var}(\widehat{\text{SVE}})/(1-\text{SVE}^2)^2$ for which 
an estimator
$\widehat{\text{Var}}(\hat u)$
can be constructed
by replacing
SVE with $\widehat{\text{SVE}}$ and
$\text{Var}(\widehat{\text{SVE}})$ with the  variance estimator in (\ref{eq:var}). The corresponding CI for SVE is obtained by back-transformation:
\begin{equation*}
    \left[\tanh\left(\widehat u -z_{1-\alpha/2}\sqrt{\widehat{\text{Var}}(\widehat u)}\right), \tanh\left(\widehat u +z_{1-\alpha/2}\sqrt{\widehat{\text{Var}}(\widehat u)}\right)\right].
\end{equation*}
Hereafter, these are termed \textit{tanh-Wald intervals}.

\subsection{Extension to Relative Effect Measures} \label{sec:relative}

In some settings a vaccine's effect may be estimated by $\widehat{\text{VE}}=1-\hat \theta$ where $\hat \theta$ is an estimated rate ratio from a Poisson regression model 
or an estimated hazard ratio from a Cox proportional hazards model.
The SVE framework extends naturally to any multiplicative relative effect measure $\theta$
by defining
\begin{equation}
    \text{SVE}=\frac{1-\theta}{\max(1,\theta)}, 
    \label{eq:sve_theta}
\end{equation}
where $\theta=1$ corresponds to the vaccine having no
effect, in which case SVE $=0$. When $\theta < 1$, the vaccine is protective and \eqref{eq:sve_theta} simplifies to $\text{SVE} = 1 - \theta$. When $\theta > 1$, the vaccine is harmful and \eqref{eq:sve_theta} becomes $\text{SVE} = (1 - \theta)/\theta$. In this setting SVE
can be estimating by replacing $\theta$
in (\ref{eq:sve_theta})
with an estimator $\hat \theta$ obtained
  using standard likelihood-based procedures, for example by the MLE from a Poisson model or the maximum partial likelihood estimator from a Cox proportional hazards models. Let $\ell(\theta)$ denote the corresponding log-likelihood or partial log-likelihood, and let $\hat\theta$ be the unconstrained maximizer of $\ell(\theta)$.

Confidence intervals
for SVE defined by \eqref{eq:sve_theta} can be constructed using  profile likelihood 
or the delta method,
similar to Sections
\ref{subsec:profile}
and
\ref{subsec:wald}.
To construct profile likelihood CIs,
for each candidate SVE value $s$, the profile likelihood $\ell_p(s)$ is obtained by maximizing $\ell(\theta)$ subject to $\text{SVE}=s$. The resulting likelihood ratio statistic $\Lambda(s)=2\{\ell(\hat\theta)-\ell_p(s)\}$ quantifies evidence against $H_0: \text{SVE}=s$. Under standard regularity conditions as $n_0,n_1\to\infty$, $\Lambda(s)$ is asymptotically $\chi^2_1$ under the null. The corresponding $(1-\alpha)$ profile likelihood confidence interval again has the form $\{s \in (-1,1):\Lambda(s)\leq\chi^2_{1,1-\alpha}\}$.

Wald-type CIs may be obtained using the delta method applied to the log-scale relative effect. Let $\hat \phi = \log(\hat \theta)$ with estimated variance $\widehat{\text{Var}}(\hat{\phi})$. The variance of $\widehat{\text{SVE}}$ can then be estimated by
\begin{equation}
    \widehat{\text{Var}}(\widehat{\text{SVE}}) = \frac{\widehat{\text{Var}}(\hat{\phi})}{\{\max(\hat\theta, \hat\theta^{-1})\}^{2}}.
    \label{eq:theta_var}
\end{equation}
When $\hat \theta=1$, (\ref{eq:theta_var}) reduces to $\widehat{\text{Var}}(\hat{\phi})$. 
\begin{comment}
As in Section~\ref{subsec:wald}, the variance estimator
(\ref{eq:theta_var})
can be unstable in small-to-moderate sample sizes. 
A variance estimator to stabilize finite-sample performance may be obtained
using the smoothing approach described above.
Here, for smoothing parameter $\varepsilon > 0$, 
the variance may be estimated by
\eqref{eq:theta_var} when $|1-\hat{\theta}| > \varepsilon$, and by
\begin{equation*}
    \widehat{\text{Var}}_\varepsilon(\widehat{\text{SVE}}) = \left(\frac{2\hat{\theta}}{1 + \hat{\theta} + \varepsilon/2}\right)^2 \widehat{\text{Var}}(\hat{\phi})
\end{equation*}
when $|1 - \hat{\theta}| \le \varepsilon$.
\end{comment}

\section{Operating Characteristics} \label{sec:sim}

This section presents results from Monte Carlo simulation studies of the SVE estimator and corresponding CIs described in Section \ref{sec:inference}. Bias, sampling variability, and CI coverage were examined across a range of infection probabilities and group sample sizes. Comparisons were also made to traditional VE-based inference  in terms of CI width and type I error rate.

\subsection{Simulation Setup}

In each simulated data set (replicate), the numbers of infected individuals in the vaccinated and unvaccinated groups were generated independently from binomial distributions with sample sizes $n_1$ and $n_0$ and infection probabilities $p_1$ and $p_0$, respectively. Underlying parameters were varied over a grid of values. Infection probabilities were set to $p_0,p_1\in\{0.1,0.2, \dots, 0.9\}$ and sample sizes $n_0, n_1\in\{100,250,1000\}$ were used. 

For each setting, $\num{10000}$ replicates were generated. For each replicate, the empirical infection risks $\hat{p}_1$ and $\hat{p}_0$ were used to compute  $\widehat{\text{SVE}}$ along with  profile likelihood, original-scale Wald, and tanh-Wald 95\% CIs. %For Wald-based CIs, the smoothing parameter $\varepsilon$ was set to $1.96\sqrt{\hat\sigma^2_0+\hat\sigma^2_1}$ to improve finite-sample coverage. Results without smoothing  are described in Web Appendix B.
Coverage probabilities were estimated by the proportions of CIs that contained the true SVE value. Type I error rates were estimated by the proportions of intervals that did not contain $0$ when the true SVE value was $0$, i.e., when $p_0=p_1$.

\subsection{Simulation Results}

\begin{figure}
    \centering
    \includegraphics[width=0.8\linewidth]{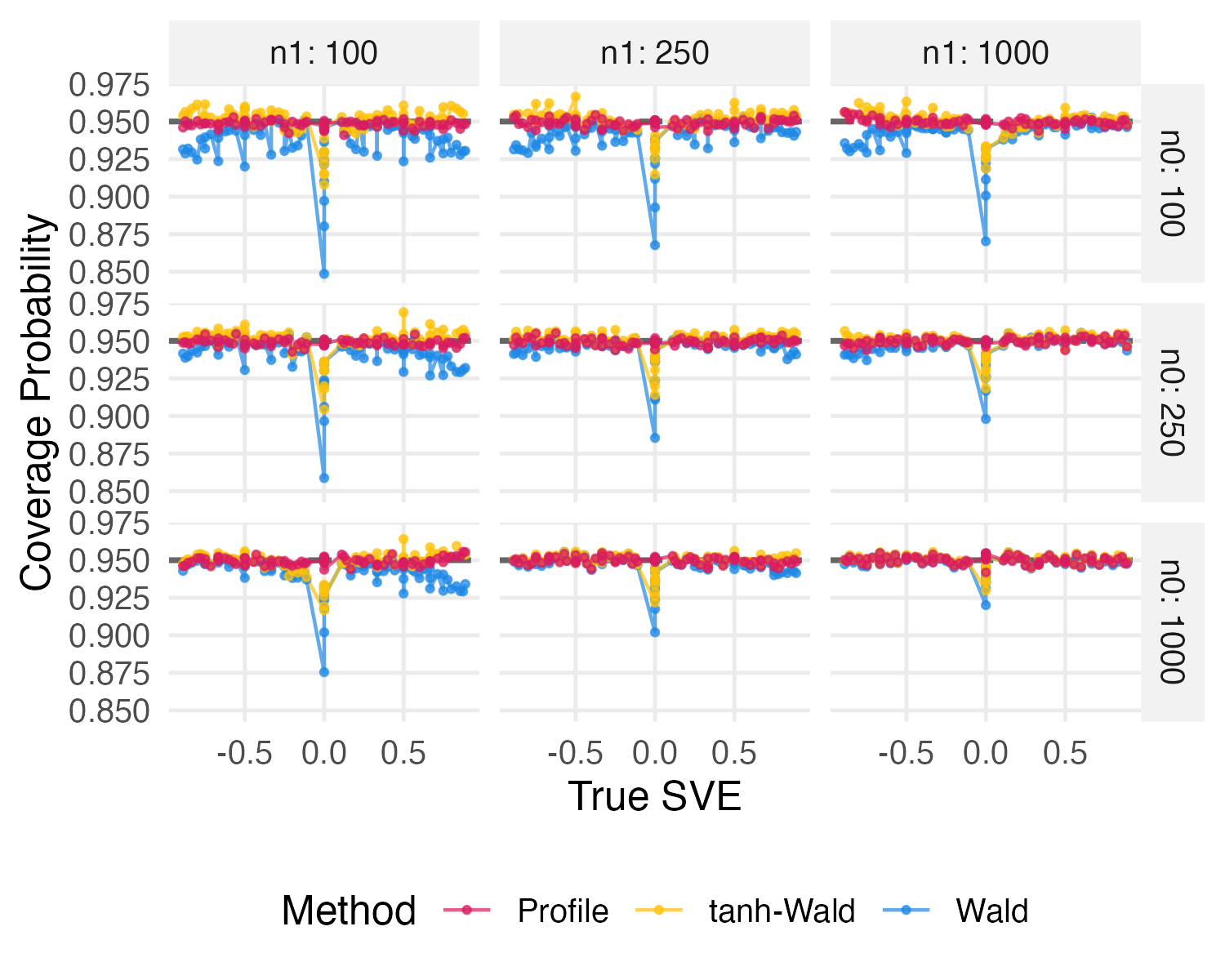}
    \caption{Empirical coverage probabilities of 95\% confidence intervals (CIs) of symmetric vaccine efficacy (SVE) as a function of true SVE values. Three types of CIs were considered: a profile likelihood interval (Profile), a Wald interval on the original SVE scale (Wald), and a back-transformed tanh-Wald interval (tanh-Wald). 
    %The Wald-based intervals include the smoothing parameter set at $\varepsilon=1.96\sqrt{\hat\sigma_0^2+\hat\sigma_1^2}$.
    Each panel corresponds to a combination of sample sizes for the vaccinated and unvaccinated groups with columns corresponding to values of $n_1 \in \{100, 250, 1000\}$ and rows to $n_0 \in \{100, 250, 1000\}$. Coverage across the full range of SVE values is close to the nominal level for the profile likelihood and tanh-Wald intervals, with modest undercoverage for the original-scale Wald intervals at smaller sample sizes. Larger sample sizes stabilize coverage for all methods.}
    \label{fig:conf}
\end{figure}

Bias of $\widehat{\text{SVE}}$ was negligible and the average  variance estimates (\ref{eq:var}) closely approximated the 
empirical variance across the range of infection probabilities and sample sizes considered (see Web Figures 1 and 2). Figure~\ref{fig:conf} displays the empirical coverage probabilities for the
 profile likelihood, original-scale Wald, and tanh-Wald CIs
as a function of true SVE, stratified by sample size. 
Across all scenarios, the profile likelihood intervals provided coverage probabilities closest to the nominal 95\% level. For small sample sizes ($n_0=n_1=100$), the profile likelihood and tanh-Wald intervals maintained coverage closer to the nominal 95\% level across the full range of SVE values, while the original-scale Wald intervals tended to undercover near extremes (i.e., when the true SVE is close to $-1$ or $1$). 
%When sample sizes were imbalanced between vaccinated and unvaccinated groups,  undercoverage of the Wald intervals depended on the true value of SVE and which group was smaller. For example, with $n_0=\num{1000}$ and $n_1=100$, the original-scale intervals undercovered for SVE close to $1$, whereas with $n_0=100$ and $n_1=\num{1000}$ the original-scale intervals undercovered when SVE was close to $-1$. In both cases, the profile likelihood and tanh-Wald intervals maintained coverage near the nominal level (Figure~\ref{fig:conf}).
As sample sizes increased, coverage for all three CIs stabilized around $95\%$. 

%In summary, both the profile likelihood and tanh-Wald intervals performed well, with the profile likelihood intervals  having near nominal coverage probabilities across all settings considered.

The empirical type I error rates for the setting when $\text{SVE}=0$ (the null case) are presented in Web Appendix B.4 (Web Figure 4). The type I error was well controlled when sample sizes were large and balanced, whereas smaller or imbalanced designs led to inflated type I errors
for the Wald intervals. The empirical type I error was closest to the nominal 5\% for the profile likelihood intervals, followed by the tanh-Wald intervals, and poorest for the original-scale Wald intervals.

\subsection{Comparison to Traditional Vaccine Efficacy}

\begin{figure}
    \centering
    \includegraphics[width=0.8\linewidth]{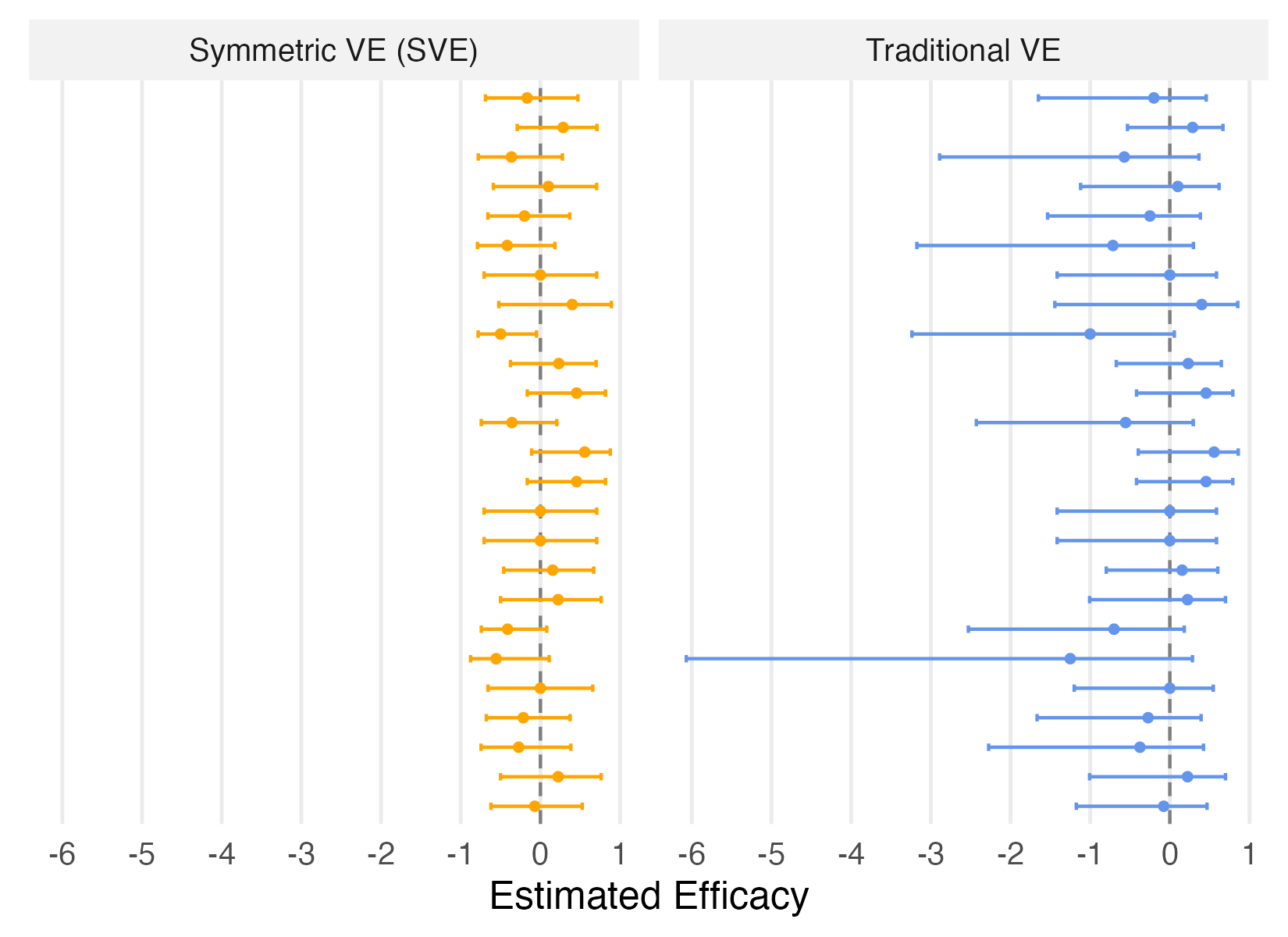}
    \caption{Forest plot of vaccine effect estimates and confidence intervals from $25$ simulated data sets under the null (no vaccine effect) with infection risk $p_0=p_1=0.1$ and sample sizes $n_0=n_1=100$. Each point represents the estimated effect with error bars representing 95\% CIs for symmetric vaccine efficacy (SVE) and traditional vaccine efficacy (VE). SVE intervals (calculated as tanh-Wald intervals) are guaranteed to be contained $[-1,1]$ and symmetrically distributed around the null, whereas traditional VE intervals are highly asymmetric, with lower bounds extending far below zero.}
    \label{fig:null}
\end{figure}

Figure \ref{fig:null} compares 
SVE 95\% CIs (tanh-Wald) to traditional VE 95\% CIs  
for $25$ randomly selected simulated data sets when $p_0=p_1=0.1$ and $n_0=n_1=100$.
The 
SVE intervals are  symmetrically distributed around the null of $0$. However, traditional VE estimates yield highly asymmetric intervals, with the lower bound often extending far below zero and potentially giving the false impression that an ineffective vaccine is actually harmful.

Additional comparison of 
the widths of 95\%  CIs for SVE and traditional VE for simulations where $n_0=n_1=\num{1000}$ are presented in Web Appendix B.3 (Web Figure 3). When the vaccine was protective (i.e., $p_1 < p_0$), estimates of SVE and VE tended to be similar and not surprisingly the corresponding CIs had similar average width. However, when the vaccine was harmful (i.e., $p_1 > p_0$), CIs for traditional VE became extremely wide, while those for SVE remained stable.  CIs for traditional VE were sometimes more than $80$ times as wide as those for SVE in the extreme cases (i.e., when $p_1$ was much higher than $p_0$).

\section{Application to an HIV Vaccine Trial} \label{sec:data}

VAX004 was a placebo-controlled phase 3 trial of a recombinant glycoprotein 120 (rgp120) vaccine designed to prevent HIV-1 infection \citep{rgp1202005placebo}. The trial randomized $\num{5403}$ participants
($2$:$1$ allocation, $n_1=\num{3598}$, $n_0=\num{1805}$)
 to receive seven injections of the rgp120 vaccine or placebo over $30$ months, with HIV-1 seroconversion as the primary endpoint.  The analysis presented here utilizes
the number of HIV-1 infections in the vaccine and placebo groups from Table~3 of \citet{rgp1202005placebo}, with VE estimated as one minus the ratio of the  proportion infected in the two trial arms. Due to the unavailability of individual-level data, the hazard ratio from a Cox proportional hazards model, as  reported in the original study, was not estimated. 

\begin{figure}[h]
    \centering
    \includegraphics[width=0.8\linewidth]{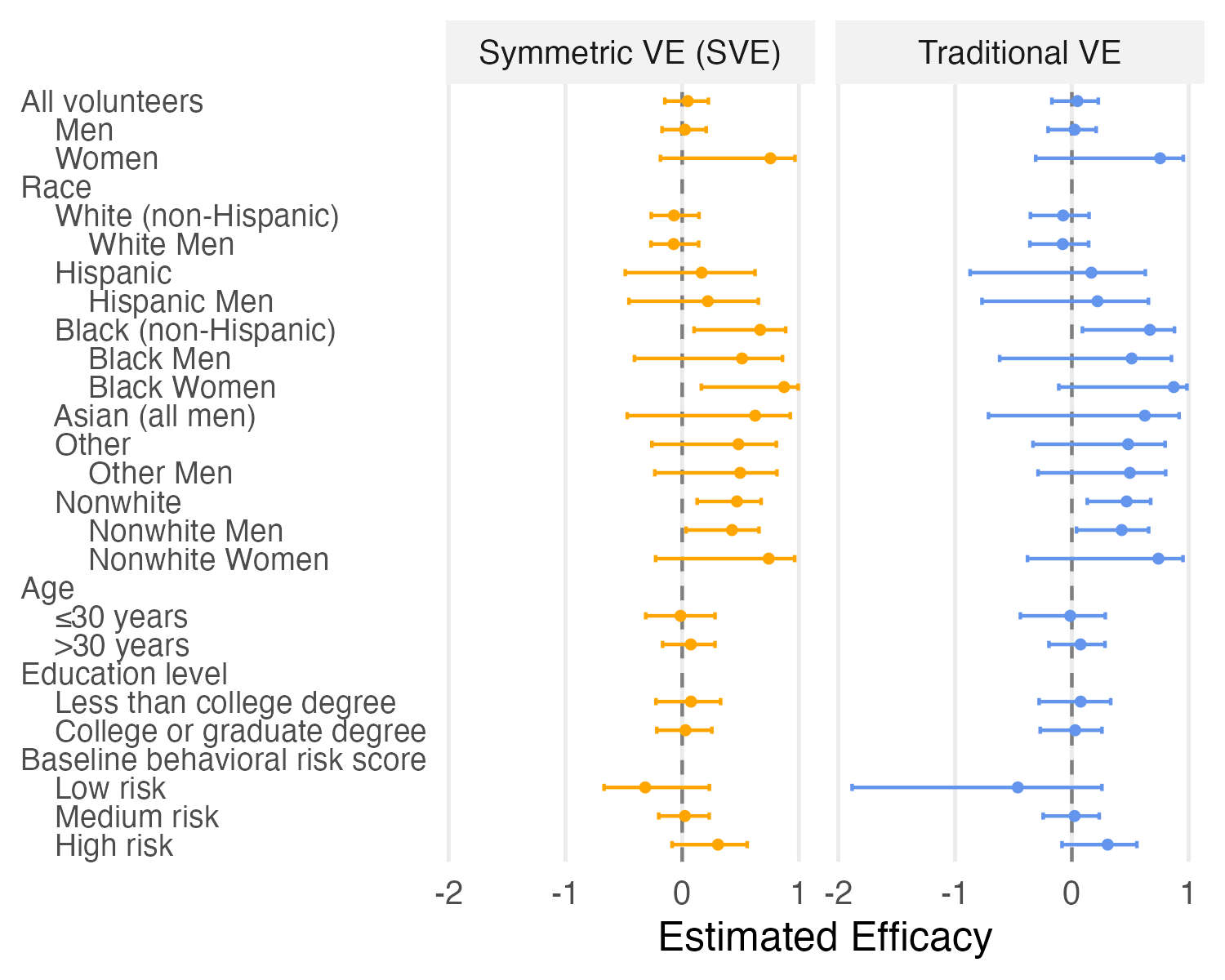}
    \caption{Forest plot comparing traditional vaccine efficacy (VE) and symmetric vaccine efficacy (SVE) against HIV-1 infection in the rgp120 Trial. Each point represents the estimated efficacy with error bars representing 95\% CIs. SVE CIs were calculated using the profile likelihood method. Data  from \citet{rgp1202005placebo}. }
    \label{fig:forest}
\end{figure}

VE was estimated overall and across several subgroups (e.g., by age, race, and baseline behavioral risk score), as in the original study. Overall and subgroup-level infection proportions as well as VE and SVE estimates are given in Web Appendix C (Web Table~1). Figure~\ref{fig:forest} presents a forest plot of $\widehat{\text{VE}}$ and $\widehat{\text{SVE}}$ with 95\% CIs across the subgroups. The 95\% CIs for the SVE were calculated using the profile likelihood method.

The results in Figure~\ref{fig:forest}
illustrate two key advantages 
of SVE
over traditional VE. First, $\widehat{\text{SVE}}$ and the corresponding profile likelihood CI  are bounded within $[-1,1]$, which precludes extremely negative point estimates or CI endpoints. This  is particularly important in smaller subgroups where there may be substantial uncertainty.
For example, in the low baseline behavioral risk score subgroup, the VE 95\% CI extends far below zero
($-1.88$ to $0.26$), whereas the SVE 95\% CI  only spans $-0.67$ to $0.23$. 

Second, as a symmetric measure, SVE is easier to interpret than traditional VE. A negative SVE value indicates a protective effect of the placebo relative to the vaccine 
on the same scale for which a positive SVE
indicates a protective effect of the vaccine relative to the placebo. For example, 
in the low baseline behavioral risk score subgroup,
the lower bound $-0.67$ of the SVE CI  indicates that the placebo group had 67\% reduced infection risk relative to the vaccine group, while the CI upper bound 0.23 
indicates that the vaccine group had 23\% reduced infection risk relative to the placebo group.
For this particular subgroup, there is considerable uncertainty due to the small sample sizes and relatively low proportion infected ($2.9\%$ in the vaccinated group versus $1.8\%$ in the unvaccinated group). As a result, the CI for traditional VE occupies a disproportionately large span in the forest plot in Figure~\ref{fig:forest}, which may give the misleading impression that the vaccine is likely to be  harmful in this subgroup. To the contrary, these results plausibly mirror the simulated results in Figure~\ref{fig:null}, where the true effect is null rather than harmful. On the other hand,  SVE provides a more balanced depiction of the possible vaccine effects across the subgroups. The bounded CIs have more reasonable width, better reflecting the underlying uncertainty without potentially implying exaggerated harm. 
That said,  
SVE and  VE are distinct parameters, and therefore CIs for SVE and VE are not directly comparable.
The narrower SVE intervals in this application reflect the bounded nature of the parameter itself, i.e., SVE is constrained by construction, whereas  VE is unbounded below. As a result, SVE CIs necessarily fall within [-1, 1], unlike VE CIs.

\section{Software Implementation in R} \label{sec:r}

To facilitate estimation of SVE and it associated uncertainty, the R package \texttt{sve} was developed. The package provides functions for estimating  SVE, the estimator's variance, and CIs as described in Section~\ref{sec:inference}. The \texttt{sve} package can be installed from GitHub using the \texttt{devtools} package:

\begin{verbatim}
# install.packages("devtools")
devtools::install_github("LucyMcGowan/sve")
\end{verbatim}

The core function \texttt{est\_sve()} outputs an  SVE estimate with a corresponding CI based on inputs consisting of the observed number of infections and sample sizes for the vaccinated and unvaccinated groups. For example, suppose in a two-arm placebo-controlled randomized vaccine trial
that the observed proportion of infected
participants is $\hat{p}_0 = 0.1$ 
in the placebo arm
and $\hat{p}_1 = 0.05$ in the vaccine arm, with $n_0 = n_1 = 1000$ participants in each arm.
This code
\begin{verbatim}
library(sve)
est_sve(x0 = 100, x1 = 50, n0 = 1000, n1 = 1000)
\end{verbatim}
 produces the following output:
\begin{verbatim}
#>   estimate lower upper level  method
#> 1     0.50  0.31  0.64  0.95 Profile
\end{verbatim}
Here, the estimated SVE is $0.50$, with 95\% CI $(0.31, 0.64)$ calculated using the profile likelihood method. The profile likelihood method is applied by default (\texttt{method = "profile"}); to use tanh-Wald intervals or original-scale Wald intervals instead, the user may select \texttt{method = "tanh-wald"} or \texttt{method = "wald"}.

The \texttt{sve\_from\_model()} function computes estimates and CIs of  SVE from estimates of a relative effect measure  (e.g., hazard ratio from a Cox model, or rate ratio from Poisson regression) as described in Section~\ref{sec:relative}. For example, a Cox-model based estimate of SVE and corresponding profile likelihood 95\% CI can be computed for the simulated trial dataset (\texttt{sim\_trial\_data}) included in the \texttt{sve} package via the following code:
\begin{verbatim}
library(survival)
fit <- coxph(Surv(time, status) ~ vaccination + age + baseline_risk, 
             data = sim_trial_data)
sve_from_model(model = fit, data = sim_trial_data, effect_name = "vaccination")
#>   estimate lower upper level  method
#> 1     0.68  0.59  0.75  0.95 Profile
\end{verbatim}

\section{Discussion}

Traditional VE is widely used and easily interpretable when the vaccine is protective and uncertainty is small (e.g., in large trials). However, VE's inherent asymmetry can create difficulties in interpretability and communication. In particular, VE is unbounded below which can lead to highly asymmetric CIs in practice. Resulting extreme negative values of the VE can misleadingly suggest that a vaccine is harmful when the true effect may simply be small or null. 

The \textit{symmetric vaccine efficacy} (SVE) proposed in this paper addresses these limitations by defining efficacy on a bounded symmetric scale from $-1$ (maximally harmful) to $1$ (maximally protective). This scale provides a natural and interpretable measure that treats protection and harm equivalently. 
Standard
 large sample frequentist inference is considered in the randomized trial setting where a binary outcome (e.g., infection) is observed for all trial participants.
SVE  can easily be adapted to accommodate other relative effect measures. For example, 
in vaccine trials where time to infection is the primary outcome, VE is commonly estimated as one minus the hazard ratio from a Cox proportional hazards model, which can be used in place of the relative risk in the SVE estimator, as described in Section~\ref{sec:relative}. To support broad use of SVE, open-source software is provided for implementing the SVE estimator in R, as described in Section~\ref{sec:r}.

In Monte Carlo simulations, the SVE estimator
and  corresponding  profile likelihood and tanh-Wald confidence intervals
performed well
across a wide range of settings. 
Under the null (when the vaccine has no effect), CIs for the SVE remain symmetric around the null, while those for traditional VE can be severely asymmetric. This feature of SVE may be particularly important for subgroup analyses and early-stage studies where small numbers of infections may lead to misleadingly wide or negative lower bounds for traditional VE.

The practical advantages of SVE are highlighted in 
the HIV vaccine trial analysis presented in
Section~\ref{sec:data}. In one subgroup, the traditional VE 95\% CI had a strongly negative lower bound, visually dominating forest plots and potentially giving the false impression that the vaccine was harmful. In contrast, SVE produced a bounded interval for all subgroups while accurately conveying the uncertainty in these estimates.

Utilization of SVE may improve communication about vaccine effectiveness to non-technical audiences, policymakers, and the public, by whom negative VE estimates can be easily misinterpreted and potentially cause disproportionate concern. Relative effect measures are inherently challenging to interpret \citep{dupont1996understanding, brown2022relative}, and the additional asymmetry of traditional VE can exacerbate misunderstandings. 
A natural question is whether SVE is more interpretable than  VE for different audiences,
including public health officials, clinicians, and patients.

Traditional VE can be challenging to interpret because harm and benefit are measured on different scales, namely $(-\infty,0)$ and $(0,1]$.  
Consequently,
positive values of VE are not directly comparable to negative values of VE. 
For example, VE = 40\% and VE = $-40$\% are not effects of comparable magnitude, despite having the same absolute value. Likewise, if a VE 95\% CI equals ($-$200\%, 67\%), then correct interpretation of the  upper and lower bounds requires recognition that a 67\% reduction in risk and a 200\% increase in risk are in fact effects of comparable magnitude (but opposite direction). 

SVE, on the other hand, does not require such nuanced understanding because benefit and harm are measured on the same scale. Nevertheless, one potential concern is that SVE’s
changing denominator adds some complexity to interpretation.
In particular, whereas positive values of SVE indicate the relative reduction in risk due to vaccination, negative values of SVE quantify the relative reduction in risk due to {\it not} receiving the vaccine.

Ultimately, whether reporting VE or SVE improves understanding of vaccine effects for different groups of individuals (e.g., subject-matter experts, public health officials, and so forth) is an important empirical question.
Further study of SVE in the broader context of health communication and evidence-based medicine is needed.

\section*{Data Availability}

Code to replicate all analyses, tables, and figures can be found on GitHub here: \url{https://github.com/LucyMcGowan/symmetric-ve}. The data referenced in Section~\ref{sec:data} were obtained from Table~3 of \citet{rgp1202005placebo} and have also been made available in our GitHub repository.

\section*{Acknowledgments}

Research reported in this publication was supported by the National Institute of Allergy and Infectious Diseases of the National Institutes of Health (NIH) under Award Number R37 AI054165. The authors thank the Associate Editor and an anonymous referee for helpful comments.

\section*{Supplementary Materials}

Web Appendices, Tables, and Figures referenced in Sections~\ref{sec:inference} -- \ref{sec:data} are available with this paper\ifarxiv. 
\else{} at the Biometrics website on Wiley Online Library.
\fi

\bibliographystyle{biom} 
\bibliography{references}

\label{lastpage}
\end{document}

\maketitle

\section{Delta Method and Variance Estimation}\label{sec:delta}

This appendix provides details regarding the delta method derivation of the asymptotic variance of $\widehat{\text{SVE}}$. 
In the main text, 
the multivariate delta method is used
to show that $\widehat{\text{SVE}}$ 
is asymptotically normal. The delta method
requires
the gradient $\boldsymbol\nabla g$ of
$g(p_0,p_1)=(p_0-p_1)/\max(p_0,p_1)$
exists and is continuous throughout the interior of the parameter space. To show this,
note $g(p_0,p_1)$ is defined piecewise according to whether $p_0>p_1$ or $p_0 < p_1$. To establish differentiability, consider the partial derivatives within each region.  
When $p_0>p_1$ 
\begin{equation*}
    \frac{\partial g}{\partial p_0}=\frac{p_1}{p_0^2},
\quad
\frac{\partial g}{\partial p_1}=-\frac{1}{p_0},
\end{equation*}
whereas for $p_0 < p_1$
\begin{equation*}
    \frac{\partial g}{\partial p_0}=\frac{1}{p_1},
\quad
\frac{\partial g}{\partial p_1}=-\frac{p_0}{p_1^2}.
\end{equation*}
Next note the partial derivatives also exist
at  points $(p,p)$ along the boundary between the regions;
in particular, 
\[
\partial g/\partial p_0 =
\lim_{h \to 0^+} \frac{g(p+h,p)-g(p,p)}{h}=
\lim_{h \to 0^-} \frac{g(p+h,p)-g(p,p)}{h}=
 \frac 1 p
\] and
\[
\partial g/\partial p_1 =
\lim_{h \to 0^+} \frac{g(p,p+h)-g(p,p)}{h}=
\lim_{h \to 0^-} \frac{g(p,p+h)-g(p,p)}{h}=
- \frac 1 p.
\]
Finally, note the partial derivatives are continuous,
including at points $(p,p)$, i.e., 
\begin{equation*}
    \lim_{(p_0, p_1)\to (p,p)}\frac{\partial g}{\partial p_0}
=\frac{1}{p}, 
\quad
\lim_{(p_0, p_1) \to (p,p)}\frac{\partial g}{\partial p_1}
=-\frac{1}{p}.
\end{equation*}
Thus, the gradient of $g(p_0,p_1)$ exists and is continuous throughout the interior of the parameter space, including along the diagonal where $p_0=p_1$.
It then follows from the delta method 
that in large samples  
\begin{equation}
   \text{Var}(\widehat{\text{SVE}}) \approx
\frac 1 n
\boldsymbol\nabla g^\top\boldsymbol\Sigma\boldsymbol\nabla g
   =
   \begin{cases}
       \dfrac{p_1^2\sigma_0^2+p_0^2\sigma_1^2}{p_0^4} & \text{ if } p_0>p_1\\[12pt]
       \dfrac{p_1^2\sigma_0^2+p_0^2\sigma_1^2}{p_1^4} & \text{ if } p_1\geq p_0
   \end{cases}
   \label{eq:var_gsve_case1}
\end{equation}
where $\sigma_j^2 = p_j (1-p_j)/n_j$ for $j=0,1.$
This variance can be consistently  estimated by 
the plug-in estimator $\widehat{\text{Var}}(\widehat{\text{SVE}})$, defined by
replacing $p_j$ and $\sigma_j$ in \eqref{eq:var_gsve_case1} with $\hat{p}_j$ and $\hat\sigma_j^2$ for $j=0,1.$

\section{}
\subsection{Estimator Finite-sample Bias}

 The estimator $\widehat{\text{SVE}}$ is consistent for SVE, but exhibits finite-sample bias, a known issue for ratio estimators  \citep{jewell1986bias}. When $p_0\neq p_1$, a second-order Taylor expansion gives the following approximations to the bias. When $p_0>p_1$, $$\text{Bias}(\widehat{\text{SVE}})=\mathbb{E}(\widehat{\text{SVE}})-\text{SVE}\approx -p_1(1-p_0)/(n_0p_0^2).$$ Note
 \[
 \left|-\frac{p_1 (1-p_0)}{n_0p_0^2}\right|
 \leq \frac{p_1 (1-p_0)}{n_0p_0^2}
 \leq \frac{1-p_0}{n_0p_0}
 \leq \frac{1}{n_0p_0}
 \]
Thus the bias will tend to be small 
except in smaller trials
where $n_0 p_0$, the expected number of events in the control arm, is small.
 Likewise,
 when $p_1>p_0$,  $$\text{Bias}(\widehat{\text{SVE}})=\mathbb{E}(\widehat{\text{SVE}})-\text{SVE}\approx p_0(1-p_1)/(n_1p_1^2),$$ which is bounded
 above by $1/(n_1 p_1)$.
\begin{comment}
  Taking the one-sided limits as $p_1\rightarrow p_0=p$ gives
\begin{equation}
    \lim_{p_1\rightarrow p^-}\text{Bias}\approx -\frac{1-p}{n_0p},\qquad
    \lim_{p_1\rightarrow p^+}\text{Bias}\approx \frac{1-p}{n_1p}.
    \label{eq:bias-lims}
\end{equation}
When the sample sizes in the two groups are equal ($n_0=n_1$) the two one-sided limits in \eqref{eq:bias-lims} are equal in magnitude and opposite sign so the bias vanishes at $p_1=p_0$ to the order shown. Overall, the leading finite-sample bias decreases at rate $1/n_0+1/n_1$, is larger when the larger of $p_0$ and $p_1$ is small, and is most noticeable when the $p_0$ and $p_1$ are close and sample sizes are small. 
\end{comment}
The results above suggest the finite-sample bias of
$\widehat{\text{SVE}}$
might be partially mitigated by using 
the bias-corrected SVE estimator 
\begin{equation}
    \widehat{\text{SVE}}_{\text{BC}} = \begin{cases}
        \widehat{\text{SVE}} + \frac{\hat{p}_1(1-\hat{p}_0)}{n_0\hat{p}_0^2} & \text{if } \hat{p}_0 > \hat{p}_1, \\
        \widehat{\text{SVE}} - \frac{\hat{p}_0(1-\hat{p}_1)}{n_1\hat{p}_1^2} & \text{if } \hat{p}_1 > \hat{p}_0, \\
        \widehat{\text{SVE}} & \text{if } \hat{p}_0 = \hat{p}_1.
    \end{cases}
    \label{eq:bias-correct}
\end{equation}
The bias corrected estimator subtracts an estimator of the bias from the original SVE estimator, using plug-in estimates $\hat{p}_0$ and $\hat{p}_1$ in place of $p_0$ and $p_1$. The correction is asymptotically negligible (vanishing at rate $O(n^{-1})$) but may reduce bias in finite samples.
\begin{comment}
particularly when sample sizes are small and event rates are low or when $p_0$ and $p_1$ are similar. When $\hat{p}_0 = \hat{p}_1$, no correction is applied, consistent with the theoretical result that bias vanishes in this symmetric case.
\end{comment}

\begin{figure}
    \centering
    \begin{subfigure}[t]{0.7\textwidth}
        \centering 
        \caption{$\widehat{\text{SVE}}$}
        \includegraphics[width=\linewidth]{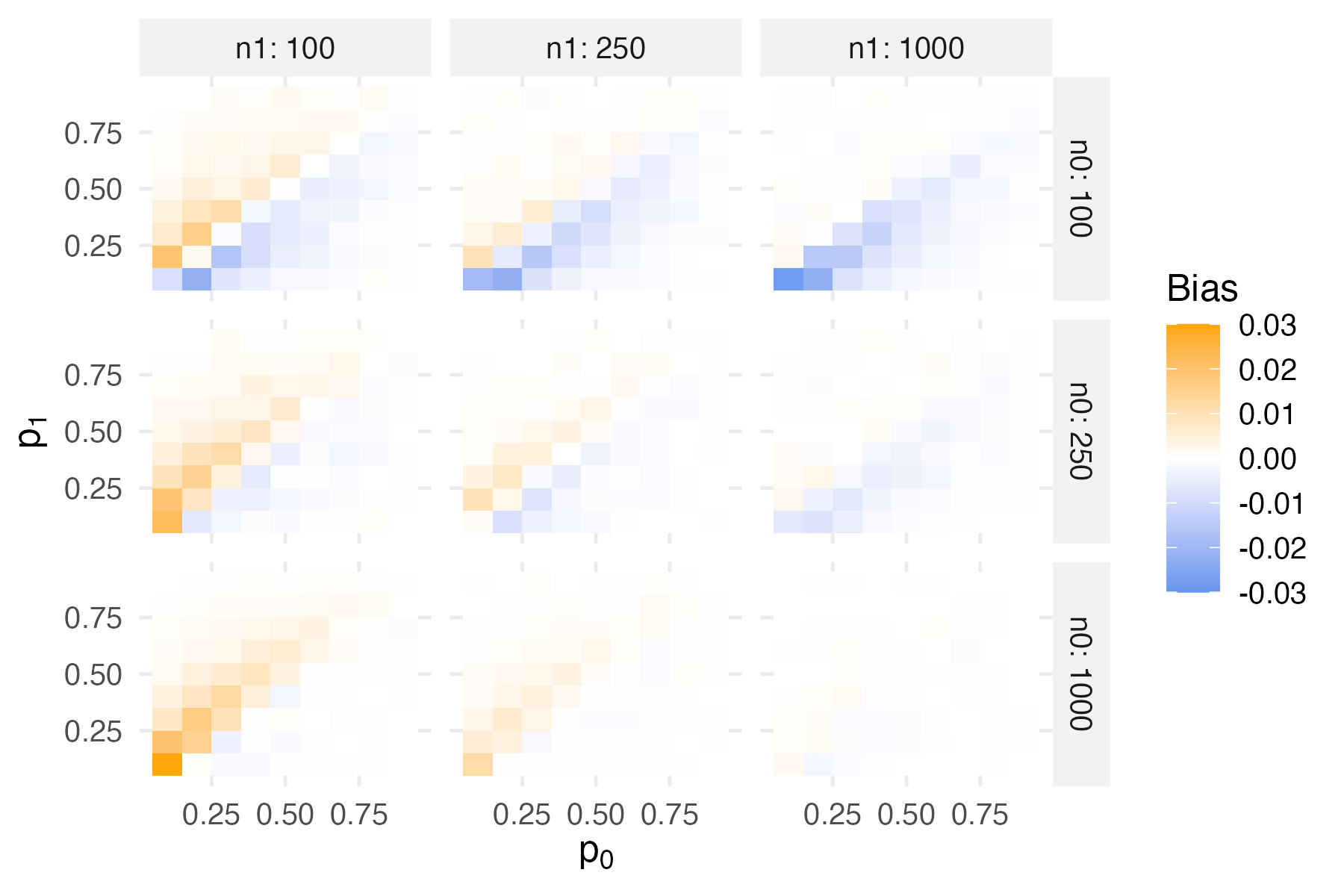}
       
        \label{fig:bias-1}
    \end{subfigure}
    \vspace{.1in}
    \begin{subfigure}[t]{0.7\textwidth}
        \centering
                \caption{$\widehat{\text{SVE}}_{\text{BC}}$}
        \includegraphics[width=\linewidth]{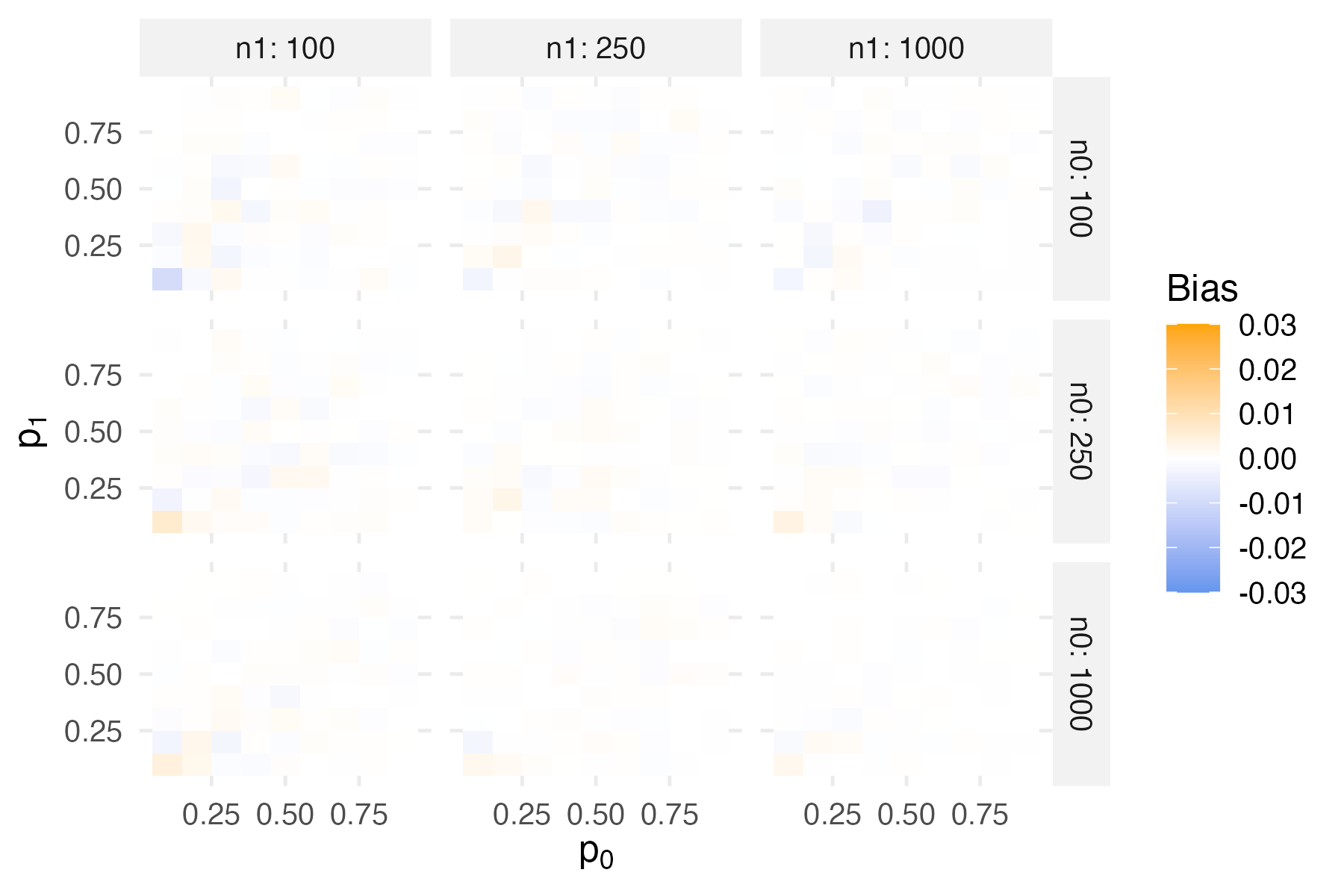}
                \label{fig:bias-corrected}
    \end{subfigure}
    \caption{Bias heatmap for the SVE estimator \textbf{(a)} without the bias correction and \textbf{(b)} with the bias correction. Each cell represents the average bias 
    ($\widehat{\text{SVE}} - \text{SVE}$ or
    $\widehat{\text{SVE}}_{\text{BC}} - \text{SVE}$) 
        over \num{10000} simulations for a given combination of risk if unvaccinated  ($p_0$), risk if vaccinated  ($p_1$) and sample sizes ($n_0$ and $n_1$). Both estimators exhibit relatively minimal bias across all scenarios, with the bias-corrected SVE estimator $\widehat{\text{SVE}}_{\text{BC}}$ having slightly reduced bias compared to the uncorrected estimator
    $\widehat{\text{SVE}}$, particularly in scenarios with smaller sample sizes and lower risks.}
    \label{fig:bias}
\end{figure}

Web Figure~\ref{fig:bias} shows the bias 
$\widehat{\text{SVE}}$ and
$\widehat{\text{SVE}}_{\text{BC}}$
across the simulation scenarios. 
The empirical bias 
of $\widehat{\text{SVE}}$ shown in
Web 
Figure~\ref{fig:bias-1} 
is small or negligible 
across all scenarios.
The finite-sample bias is greatest when
$p_0$ and $p_1$ are small and the corresponding sample size is small. 
Web Figure~\ref{fig:bias-corrected} shows the bias of $\widehat{\text{SVE}}_{\text{BC}}$, demonstrating that the bias correction effectively reduces the finite-sample bias in scenarios with small sample sizes and low event probabilities.

\subsection{Standard Error Ratio}

\begin{figure}
    \centering
    \includegraphics[width=0.9\linewidth]{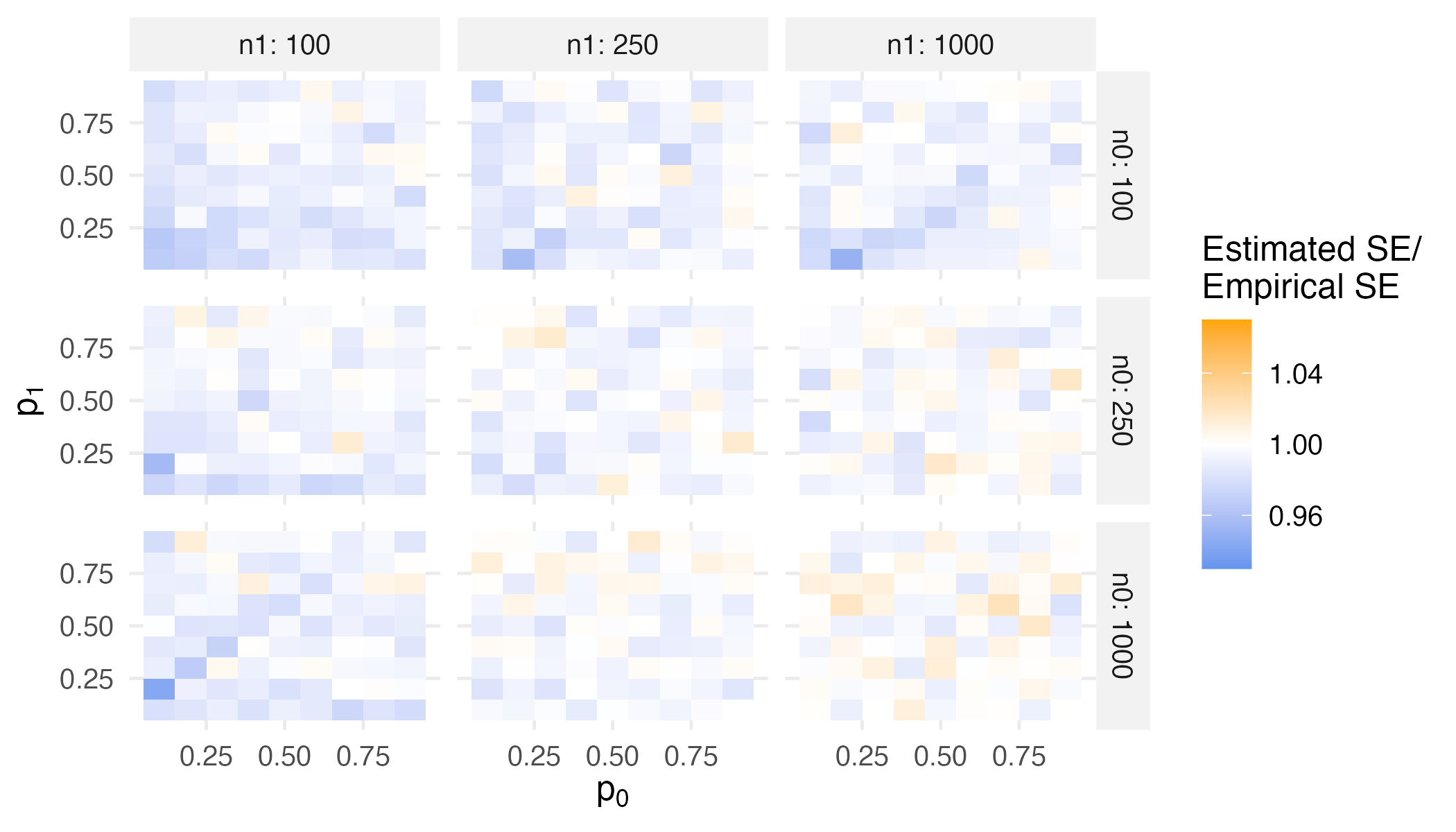}   
    \caption{Ratio of averaged estimated standard error $\widehat{\text{Var}}(\widehat{\text{SVE}})^{1/2}$ to the empirical standard error (SE) of $\widehat{\text{SVE}}$ from the simulation study. Values near 1 (white) indicate accurate variance estimation. Orange indicates overestimation of the standard error, while blue indicates underestimation.}
    \label{fig:se_ratio}
\end{figure}

Web Figure~\ref{fig:se_ratio}
shows the ratio of averaged estimated standard error $\widehat{\text{Var}}(\widehat{\text{SVE}})^{1/2}$ to the empirical standard error (SE) of $\widehat{\text{SVE}}$ from the simulation study. The ratios are close to 1 
 across all scenarios, indicating accurate variance estimation.

\subsection{Comparison of SVE and VE Confidence Interval Widths}

\begin{figure}
    \centering
    \includegraphics[width=0.8\linewidth]{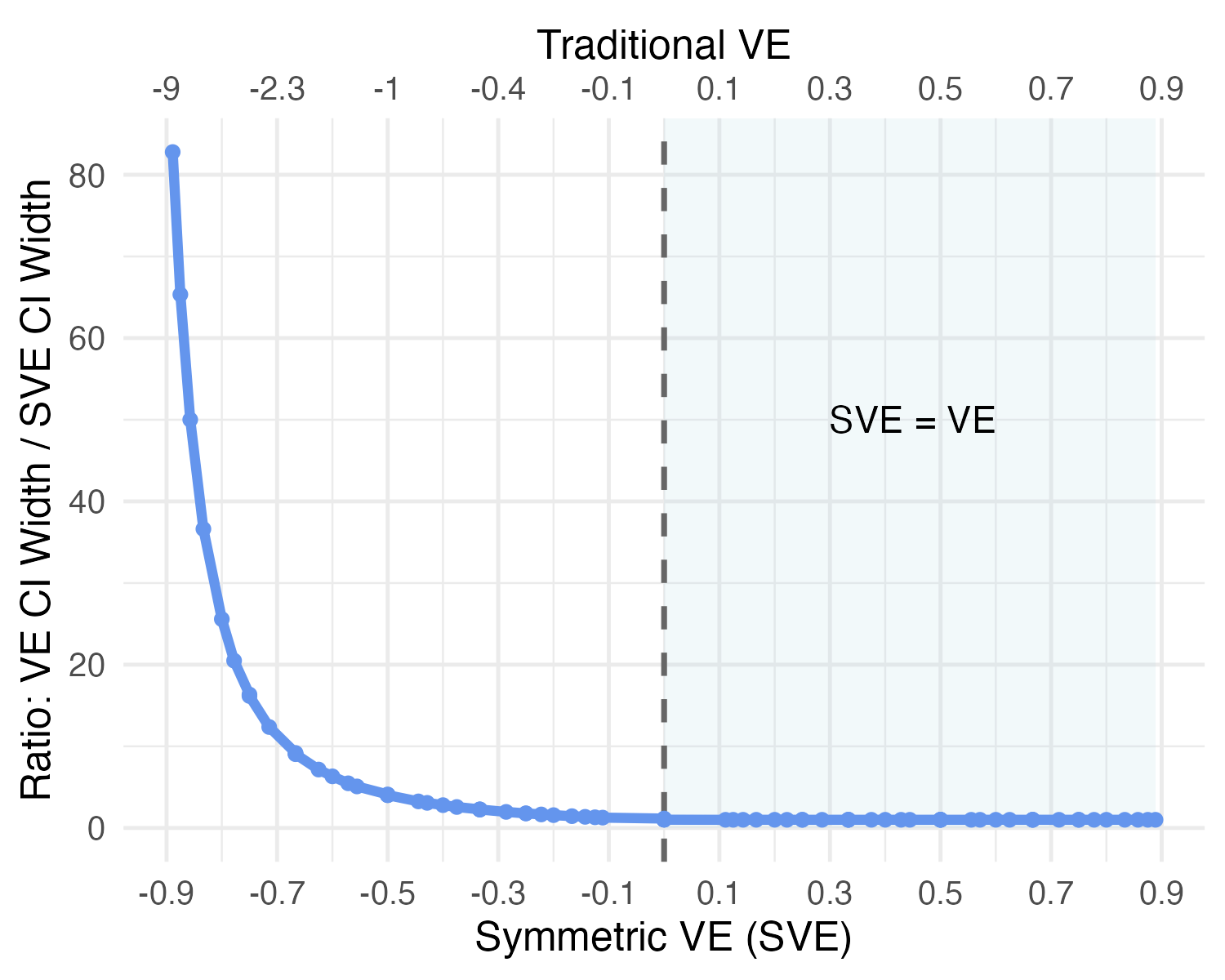}
    \caption{The ratio of 95\% confidence interval widths for traditional vaccine efficacy (VE) to symmetric vaccine efficacy (SVE).
    The bottom horizontal axis shows values of SVE, and the corresponding values of traditional VE are shown on the top horizontal axis. The shaded region indicates where $\text{VE} = \text{SVE}$, corresponding to positive efficacy. When efficacy is negative (unshaded region), the traditional VE confidence intervals can become extremely wide relative to SVE confidence intervals. The vertical dashed line represents where VE and SVE are both equal to $0$.}
    \label{fig:conf-width}
\end{figure}

Web Figure~\ref{fig:conf-width} shows the ratio of 95\% CI widths for traditional VE versus SVE for simulations where $n_0=n_1=1000$. When the vaccine is protective (i.e., $p_1 < p_0$), estimates of SVE and VE tend to be similar with CIs of similar average width. However, when the vaccine is harmful (i.e., $p_1 > p_0$), 
confidence intervals for traditional VE become extremely wide, while those for SVE remain stable. Confidence intervals for traditional VE were more than $80$ times as wide as those for SVE in the extreme case when $p_1$ was much higher than $p_0$.

\subsection{Type I Error}

Web Figure~\ref{fig:null-all} presents empirical type I error rates under the null ($\text{SVE}=0$). Type I error control was strongest for the profile likelihood intervals, followed by the tanh-Wald intervals, and poorest for the original-scale Wald intervals.

\section{Application to an HIV Vaccine Trial} 

 Overall and subgroup-level 
 empirical 
 HIV-1 infection risks 
 are shown in Web Table~\ref{tab:study},
 as well as VE and SVE estimates with corresponding 95\% confidence intervals (CIs). The CIs for  SVE were obtained using the profile likelihood method.

\bibliographystyle{biom} 
\bibliography{references}

\newpage
\pagebreak[4]

\begin{table}[H]

\caption{Estimated risk of HIV-1 infection (number of infections/number of participants) in the vaccine and placebo groups, with estimates and 95\% confidence intervals (CIs) of traditional vaccine efficacy (VE) and symmetric vaccine efficacy (SVE)}\label{tab:study}
\centering
\resizebox{\columnwidth}{!}{
\begin{tabular}[t]{lcccc}

\toprule
Category & Vaccine & Placebo & VE (95\% CI) & SVE (95\% CI)\\
\midrule
All volunteers & 241/3598 (0.07) & 127/1805 (0.07) & 0.05 (-0.17 to 0.23) & 0.05 (-0.15 to 0.22)\\
\quad Men & 239/3391 (0.07) & 123/1704 (0.07) & 0.02 (-0.20 to 0.21) & 0.02 (-0.17 to 0.21)\\
\quad Women & 2/207 (0.01) & 4/101 (0.04) & 0.76 (-0.31 to 0.95) & 0.76 (-0.19 to 0.97)\\
Race &  &  &  & \\
\quad White (non-Hispanic) & 211/2994 (0.07) & 98/1495 (0.07) & -0.08 (-0.36 to 0.15) & -0.07 (-0.26 to 0.14)\\
%\addlinespace
\qquad Men & 211/2930 (0.07) & 98/1468 (0.07) & -0.08 (-0.36 to 0.14) & -0.07 (-0.27 to 0.14)\\
\quad Hispanic & 14/239 (0.06) & 9/128 (0.07) & 0.17 (-0.87 to 0.63) & 0.17 (-0.49 to 0.62)\\
\qquad Men & 13/211 (0.06) & 9/114 (0.08) & 0.22 (-0.77 to 0.66) & 0.22 (-0.46 to 0.65)\\
\quad Black (non-Hispanic) & 6/233 (0.03) & 9/116 (0.08) & 0.67 (0.09 to 0.88) & 0.67 (0.10 to 0.89)\\
\qquad Men & 5/121 (0.04) & 5/59 (0.08) & 0.51 (-0.62 to 0.85) & 0.51 (-0.41 to 0.86)\\
%\addlinespace
\qquad Women & 1/112 (0.01) & 4/57 (0.07) & 0.87 (-0.11 to 0.99) & 0.87 (0.16 to 0.99)\\
\quad Asian (all men) & 3/56 (0.05) & 3/21 (0.14) & 0.62 (-0.71 to 0.92) & 0.62 (-0.47 to 0.93)\\
\quad Other & 7/76 (0.09) & 8/45 (0.18) & 0.48 (-0.33 to 0.80) & 0.48 (-0.26 to 0.81)\\
\qquad Men & 7/73 (0.10) & 8/42 (0.19) & 0.50 (-0.29 to 0.80) & 0.50 (-0.23 to 0.81)\\
\quad Nonwhite & 30/604 (0.05) & 29/310 (0.09) & 0.47 (0.13 to 0.68) & 0.47 (0.13 to 0.68)\\
%\addlinespace
\qquad Men & 28/461 (0.06) & 25/236 (0.11) & 0.43 (0.04 to 0.66) & 0.43 (0.03 to 0.66)\\
\qquad Women & 2/143 (0.01) & 4/74 (0.05) & 0.74 (-0.38 to 0.95) & 0.74 (-0.23 to 0.96)\\
Age &  &  &  & \\
\quad $\leq30$ years & 84/971 (0.09) & 43/504 (0.09) & -0.01 (-0.44 to 0.29) & -0.01 (-0.31 to 0.28)\\
\quad $> 30$ years & 157/2627 (0.06) & 84/1301 (0.06) & 0.07 (-0.20 to 0.28) & 0.07 (-0.17 to 0.28)\\
%\addlinespace
Education level &  &  &  & \\
\quad Less than college degree & 95/1409 (0.07) & 52/713 (0.07) & 0.08 (-0.28 to 0.33) & 0.08 (-0.22 to 0.33)\\
\quad College or graduate degree & 146/2188 (0.07) & 75/1092 (0.07) & 0.03 (-0.27 to 0.26) & 0.03 (-0.22 to 0.25)\\
Baseline behavioral risk score &  &  &  & \\
\quad Low risk & 32/1211 (0.03) & 11/609 (0.02) & -0.46 (-1.88 to 0.26) & -0.32 (-0.67 to 0.23)\\
%\addlinespace
\quad Medium risk & 177/2229 (0.08) & 90/1107 (0.08) & 0.02 (-0.25 to 0.23) & 0.02 (-0.20 to 0.23)\\
\quad High risk & 32/158 (0.20) & 26/89 (0.29) & 0.31 (-0.08 to 0.56) & 0.31 (-0.09 to 0.56)\\
\bottomrule
\end{tabular}
}
\end{table}

\clearpage

\clearpage

\clearpage

\clearpage

\begin{figure}
    \centering
    \includegraphics[width=0.8\linewidth]{figures/fig-s4.png}
    \caption{Type I error rates under the null hypothesis (vaccine has no effect) across varying shared infection risks ($p_0 = p_1$) and sample sizes ($n_0, n_1$). Three types of confidence intervals were considered: a profile likelihood interval (Profile), a Wald interval on the original SVE scale (Wald), and a back-transformed tanh-Wald interval (tanh-Wald). Each panel displays the empirical type I error as a function of the infection risk, with columns corresponding to the vaccinated group sample size $n_1$ and rows to the unvaccinated group sample size $n_0$. The dashed horizontal line marks the nominal level of $0.05$.  When sample sizes are large and balanced (e.g., $n_0 = n_1 = \num{1000}$), type I error is relatively well controlled across all infection risks. However, type I error is more variable in smaller sample sizes, particularly at small infection risks (e.g., $p_0 = p_1 = 0.1$), where noticeable inflation is observed for both Wald-type methods. The profile likelihood interval maintains nominal type I error across all scenarios.
}
    \label{fig:null-all}
\end{figure}

\label{lastpage}